\documentclass[useAMS,usenatbib]{mn2e}

\input psfig.sty

\def\aj{AJ}%
\def\actaa{Acta Astron.}%
\def\apj{ApJ}%
\def\aap{A\&A}%
\def\mnras{MNRAS}%
\def\pasp{PASP}%
\def\nat{Nature}%
\begin{document}

\title[LMC star clusters]{Disentangling the physical reality of star cluster candidates 
projected towards the inner disc of the Large Magellanic Cloud}

\author[A.E. Piatti]{Andr\'es E. Piatti$^{1,2}$\thanks{E-mail: 
andres@oac.uncor.edu}\\
$^1$Observatorio Astron\'omico, Universidad Nacional de C\'ordoba, Laprida 854, 5000, 
C\'ordoba, Argentina\\
$^2$Consejo Nacional de Investigaciones Cient\'{\i}ficas y T\'ecnicas, Av. Rivadavia 1917, C1033AAJ,
Buenos Aires, Argentina \\
}

\maketitle

\begin{abstract}

We have used Washington photometry for 90 star cluster candidates of small angular size
-typically $\sim$ 11$\arcsec$ in radius- distributed within nine selected regions in the
inner disc of the Large Magellanic Cloud (LMC) to disentangle whether they are
genuine physical system, and to estimate the ages for the confirmed clusters.
In order to avoid a misleading interpretation of the cluster colour-magnitude
diagrams (CMDs), we applied a subtraction procedure to statistically clean them from field star 
contamination. Out of the 90 candidate clusters studied, 61 of them resulted to be genuine 
physical systems, whereas the remaining ones were classified as possible non-clusters 
since either their CMDs and/or
the distribution of stars in the respective fields do not resemble those of stellar
aggregates. We statistically show that $\sim$ (13 $\pm$ 6)$\%$ of the catalogued clusters 
in the inner disc could be possible non-clusters, independently of their deprojected distances. 
We derived the ages for the confirmed clusters from the fit of theoretical
isochrones to the cleaned cluster CMDs. The derived ages resulted to be in the age range 
7.8 $\le$ log($t$) $\le$ 9.2. Finally, we built cluster frequencies for the different studied 
regions and found that
there exists some spatial variation of the LMC CF throughout the inner disc. 
Particularly, the innermost field contains a handful of clusters 
older than $\sim$ 2 Gyr, while the wider spread between different
CFs has taken place during the most recent 50 Myr of the galaxy lifetime.

\end{abstract}


\section{Introduction}

Star clusters have long been key objects to reconstruct the formation and the 
dynamical and chemical evolutions of galaxies. As the Large Magellanic Cloud (LMC) 
is considered, the study of its star cluster population has allowed us to learn
about its spread in metallicity at the very early epoch \citep{betal96}; the 
existence of a relatively important gap in its age distribution \citep{getal97}; 
the evidence of vigorous star cluster formation episodes \citep{p11a}; the complexity 
of the cluster formation rate during the last million years \citep{dgetal13}, etc. 
Although somewhat inhomogeneous and clearly still incomplete, the astrophysical 
properties estimated for a significant number of LMC star clusters have been the 
starting point to address galactic global issues such as the age-metallicity 
relationship and the cluster formation rate, among others. Therefore, it results
of great importance to enlarge the number of objects confirmed as genuine star
clusters and to estimate their fundamental parameters.

With the aim of providing mainly with age and metallicity estimates for an 
increasing number of LMC star clusters, we have continued a long term observational 
program carried out at Cerro Tololo Interamerican Observatory (CTIO) using different
telescopes in conjunction with  CCD cameras and the Washington photometric filters 
\citep{c76}. A total of 61 clusters have been observed and their fundamental 
parameters estimated 
\citep[see, e.g.][]{getal97,petal99,petal02,getal03,petal03a,petal03b,petal09,petal11}.
More recently, we took advantage of a wealth of available images at the
National Optical Astronomy Observatory (NOAO) Science Data Management
(SDM) Archives\footnote{http://www.noao.edu/sdm/archives.php}, obtained at the
CTIO 4-m Blanco telescope with the Mosaic II camera attached (36$\times$36 arcmin$^2$ 
field with a 8K$\times$8K CCD detector array, scale 0.274$\arcsec$/pixel) and the Washington filters. From the 
whole volume of observed LMC fields \citep[see][]{petal12} we identified 206
clusters previously catalogued by \citet[hereafter B08]{betal08}, and studied 107 of them 
with some detail \citep{p11a,p12a}.
In this work, we end up the series of studies of unkown or poorly-known LMC clusters 
with available Washington photometry by analysing the remaining objects in
the aforementioned sample.

The paper is organised as follows: Washington $C,T_1$ data are presented in Section 2.
 The star cluster sample is described in Section 3, while the cleaning of the colour-magnitude diagrams
 (CMDs) and the estimation of the cluster ages are presented in Sections 
4 and 5, respectively. We discuss the results and build cluster frequencies in Section 6.
Finally, conclusions of this analysis are given in Section 7.

\section{Data handling}

The Washington $C,T_1$ photometric data set used in this work were obtained from an 
comprehensive process that involved the reduction of the raw images, the determination of the 
instrumental photometric magnitudes, and the standardisation of the photometry. We have 
already described in detail such steps not only in the works cited above, but also in 
\cite{p11b,p11c,p12b,pb12,metal14}. For this reason, we summarise here some specific 
issues in order to provide the reader with an overview of the photometry quality.

The data come from observations carried out on the nights of Dec. 18-20, 2008
through the CTIO 2008B-0192 program (PI: D. Geisler). The nights resulted to be of excellent
photometric quality  -with a typical seeing of $\sim$ 1$\arcsec$ \citep[see Table 1 of][]{petal12}-
as judged by the rms errors from 
the transformation to the  standard system ($\sim$ 0.02 mag in each filter). From
extensive artificial star tests we showed that the 50$\%$ completeness level is located at 
$C$ $\sim$ 23.5-24.5 mag and $T_1$ $\sim$ 23.0-24.0 mag, depending on the crowding and exposure 
 time \citep[see][]{petal12}.
%We only analysed data here for cluster fields
%to the magnitude where the completeness level begins to fall below 100$\%$. 
The photometric
errors in the $T_1$ mag increase typically from $\sigma(T_1)$ $\le$ 0.01 mag for the $T_1$ 
range $\sim$ 16.0-20.0 mags up to $\sigma(T_1)$ $\sim$ 0.20 mag for $T_1$ = 24.0 mag. The 
errors for the $C-T_1$ mags resulted in $\sigma(C-T_1)$ $\le$ 0.01 for  the $T_1$ range $\sim$ 
16.0-19.0 mags and increases exponentially up to $\sigma(T_1)$ $\sim$ 0.30 mag for $T_1$ = 24.0 mag.
We here analyse cluster CMDs down to $T_1$ $\sim$ 22.0 mag, which corresponds to a completeness
level nearly 100\% and photometric errors $\sigma(T_1)$ $\le$ 0.05 mag. Thus, our
photometry yield accurate morphology and position of the main features
in the observed CMDs of the studied fields.

\section{Defining the cluster sample}

Fig. 1 shows a schematic chart with the distribution of the cluster candidates catalogued 
by B08 (dots) and the positions of the Mosaic II fields wherein the objects studied 
here are located (boxes). Some of the clusters placed
within these Mosaic II fields have been previously studied by different authors
 \citep{betal13,p11a,p12a,chetal14}, while others have been discarded from this analysis since their photometric
data are not complete due to the presence of saturated stars in their images (BSDL\,351, 349, 1980, H88-267, 
308, KMHK\,1274, HS\,81, OGLE\,308, and SL\,360). The final cluster candidate 
sample studied here contains 90 objects.

According to B08's catalogue the objects under study are of small angular
dimension, typically with radius between $\sim$ 8$\arcsec$ and 19$\arcsec$ with an average of 
11$\arcsec$. Considering the LMC regions traced
by \cite{hz09}, some clusters appear projected towards the densest 30 Doradus, the Bar, and the 
outer Bar regions, while others are distributed within the North-west Void and the North-west
Arm, which are less crowded. All these substructures are within the so-called inner LMC disc 
\citep[R $<$ 4 kpc; ][]{betal98}. In addition, many of the tiny selected objects are apparently composed by a handful 
of stars, which makes them of very low contrast over the background stellar density.

The clusters included in this study were first recognised by overplotting the positions of 
those catalogued by B08 to the deepest Mosaic II $C, T_1$ images, thus avoiding to mismatch 
the  observed objects and the actual list of catalogued clusters. We thus assigned 
to the observed clusters the respective names taken from B08. We searched for the same names
 in the  Digital Sky Survey 
(DSS)\footnote{The Digitized Sky Surveys were produced at the Space Telescope Science Institute 
under U.S. Government grant NAG W-2166. The images of these surveys are based on photographic 
data obtained using the Oschin Schmidt Telescope on Palomar Mountain and the UK Schmidt Telescope. 
The plates were processed into the present compressed digital form with the permission of these 
institutions.} and downloaded 15$\arcmin$$\times$15$\arcmin$ $B$ images centred on the coordinates 
matched by the DSS. We used the SIMBAD Astronomical Database as an additional source for checking the
cluster coordinates. When comparing the DSS extracted regions with the observed cluster 
fields in the $C,T_1$ images, we could confirm the positions of our selected targets.
Notice that, since most of the observed objects are of small angular size and many of them
are projected towards relatively crowded fields with stellar density fluctuations, 
the task of distinguishing a star cluster from a chance grouping of stars was not at all
straightforward. Table 1 contains a complete list of the recognized objects. 

\section{Analysis of the Colour-Magnitude Diagrams}

In general terms, the observed CMDs of the selected objects are the result of the superposition
of different stellar populations distributed along the line-of-sight. For this reason, the
use of the observed CMDs without subtracting the luminosity function and the colour distribution 
of stars belonging to the field might lead to wrong interpretations. Moreover, since the
catalogued clusters have been identified as small star concentrations on the basis of a stellar 
density fluctuation in the sky, their real physical nature require a subsequent confirmation.
In most of the cases we probably deal with the presence of a genuine star cluster, while
in other cases it might constitute a chance grouping of stars or the effect
of a non-uniform distribution of the interstellar material in that surveyed region. 
In order to disentangle cluster stars from field stars, it is required
to employ CMDs of adjacent fields to subtract the local LMC field luminosity function and colour 
distribution. 

We built cluster CMDs from all the measured stars distributed within a circle
of radius three times those of the clusters. For this purpose we used the radii (r) listed
in Table 1, which were obtained from visual inspection of the objects 
in the deepest $C,T_1$ images. They are large enough as to reach the observed 
field star region. Notice that our main aim consists in cleaning the cluster CMDs from
the contamination of field stars within areas  around the clusters' centres nine times
larger than $\pi$r$^2$, so that we did not need to trace their radial profiles.  
Once the cluster areas were delineated, we traced four additional
regions with areas equals to the cluster regions and placed more or less equidistant to the
clusters' centres 
around the cluster circular areas. From each surrounding region we built a CMD which
shows the features of the local LMC star field towards that particular direction.
We then apply a procedure that compares each one of the field CMDs to the cluster CMD
and subtract from the latter a representative field CMD in terms of stellar density, 
luminosity function and colour distribution. We refer the reader to \cite{pb12} for details 
concerning this decontamination method. Each resultant representative 
field CMD is built from scanning the individual one using cells that vary in size, thus
achieving a better field representation than counting the number of stars in boxes fixed 
in size. Then, the stars in the cluster CMD that fall within the defined cells and
closest to the representative positions are eliminated. 

The method allows that cells vary in magnitude and colour separately according to the free
path between field stars in the CMD, so that they result bigger in CMD regions with a 
small number of stars, and vice versa.  The free path is defined as
$\Delta$(colour)$^2$ + $\Delta$(magnitude)$^2$  = (free path)$^2$, 
where $\Delta$(colour) and $\Delta$(magnitude) are the distances from the considered star to the 
closest one in abscissa and ordinate in the field CMD. In practice, an initial rectangular 
cell with a dimension of ($\Delta$(colour), $\Delta$(magnitude) = (0.5, 1.0) is put on a single
field star. Then, the method looks for the closest star in magnitude and colour (it will be
placed at a corner of the rectangular cell), so that the resulting closest magnitude and colour
will be used to define the free path of the considered star. The task is 
repeated for every star in the field CMD. Therefore, each field star has associated a different 
cell. These cells are then superimposed to the cluster CMD, and the stars closest to their centres
are eliminated. In this sense, the cell sizes depend not 
only on the stellar density of that field (the denser a field the more stars in the field CMD), 
but also on the magnitude and colour distributions
of those field stars in the CMD, making some parts of the field CMD  more populated than 
others. For instance, relatively bright field red giants with small photometric errors
usually appear relatively isolated at the top-right zone of the CMD, while faint Main
Sequence (MS) stars are more numerous at the bottom part of the CMD. For this reason, 
bigger cells are required to satisfactorily subtract stars from the cluster CMD regions
where there is a small number of field stars, while smaller cells are necessary for 
those CMD regions more populated by field stars.

After repeating the procedure of subtracting stars from the cluster CMD using the four 
selected surrounding fields, we obtained four cleaned cluster CMDs which were compared to 
each other. As a general rule, each star in the observed cluster CMD keeps not subtracted 
different times; the more the times a star is subtracted, the higher the probability 
of being a field feature. Thus, we convert the number of times that a star appears in the 
four cleaned cluster CMDs
in an estimation of its probability (P) of being a fiducial feature in that cluster CMD.
For instance, stars that are considered to belong to the field population appear once, or
do not appear in the four cleaned cluster CMDs (we refer to them as of P $\le$ 25\%); stars
that could indistinguishably belong to the field or to the studied object are seen twice
(P = 50$\%$); and stars that are predominatly found in the cleaned cluster area
rather than in the star field population  (P $\ge$ 75\%) are identified more than twice.
Whenever the CMDs for stars with P $\le$ 25\%
looks like that of P $\ge$ 75\%, we conclude that both the star field and the
cluster are of similar features (ages).

To illustrate the bulk of the analysis performed, we produced multiple panel figures
for the objects listed in Table 1. As an example, we include Fig. 2, which depicts three
CMDs and an enlargement of the $T_1$ image centred on SL\,390. We overplotted to the
enlarged image a circle representing the cluster radius used in the analysis and marked
the stars that have a chance of being cluster members P $\ge$ 75$\%$. The three
CMDs represent the observed cluster CMD for the stars measured within the cluster radius 
(upper-left panel); a single field CMD for an annulus centred on the cluster, with an internal 
radius 3 times that of the cluster and an equal cluster area (upper-right panel); 
and the cleaned cluster CMD (bottom-left). The colour scale represents stars that 
statistically belong to the field (P $\le$ 25\%, white), stars that might belong either 
to the field or to the cluster (P  $=$ 50\%, gray), and stars that predominantly populate the 
cluster region (P $\ge$ 75\%, black). The multiple panel figures for the entire sample of 
clusters can be found at the on-line version of the Journal as Supplementary material.

The distribution of stars with different membership probabilities in the CMD as well as in the
cluster field were used at a time to confirm the physical reality of the studied cluster
candidates. We required that the stars with P $\ge$ 75\% within the cluster radius have as a 
counterpart a CMD resembling that of a star cluster to conclude that we are dealing with a 
genuine physical system. In some cases, we reinforced our conclusion about the nature of an object
from the analysis of the distribution of stars with P = 50$\%$ and $\le$ 25$\%$, respectively.
We found that a handful of objects could be possible non-clusters, since either the CMDs and/or
the distribution of stars in the respective fields do not resemble that of an stellar
aggregate (BSDL\,218, 256, 616, 661, 1103, 2794, 2824, 2841, 2842, 2891, 2898, BRHT\,60a, 62b, 
GKK-O155, 203, 205, H88-248, 318, KMHK\,125, 237, 258, 289, 609, 897, OGLE\,291, 303, 335, 344,
and SL\,371). Fig. 3 depicts the results for BSDL\,616, where the cleaned CMD shows a MS and a 
Red Clump (RC) mostly composed by field stars. Other possible non-clusters resulted to be single 
bright stars surrounded by unresolved faint ones. Fortunately, the Mosaic II images have a better 
resolution and a fainter magnitude limit than those used for catalogued cluster candidates
\citep[see details in][]{pb12}, so apparent compact extended objects could be resolved. 
As an example, Fig. 4 compares an enlargement of the $T_1$ image centred on KMHK\,125 
to that obtained from the DSS.

\section{Cluster age estimates}

We estimated ages for the 61 confirmed star clusters by using the set of theoretical isochrones
computed by \cite{metal08} for the Washington photometric system.
The isochrones were first reddened by the appropriate cluster colour excesses and corrected 
due to the distance effect. Then, we superimposed those for a metallicity level of Z = 0.008 to 
the cluster CMDs and chose the ones which best reproduced the cluster features to assign the 
cluster ages. The matching procedure was performed using stars with P $\ge$ 75$\%$.

The estimation of cluster reddening values was made by interpolating the extinction maps of 
\cite[hereafter BH]{bh82}. BH maps were obtained from H\,I (21 cm) emission data for the 
southern sky. They furnish us with foreground $E(B-V)$ colour excesses which depend on the 
Galactic coordinates. We also took advantages of the Magellanic Clouds extinction values based 
on RC stars photometry provided by the OGLE collaboration \citep{u03} as described
in \citet{hetal11}. They resulted in average (0.02 $\pm$ 0.01) mag smaller than those obtained 
by BH. As for the \citet[hereafter SFD]{setal98} full-sky maps from 100-$\mu$m dust emission, 
we decided not to use them since the authors found that deviations are coherent in the sky and 
are especially conspicuous in regions of saturation of H\,I emission towards denser clouds and 
of formation of H$_2$ in molecular clouds. The relatively small angular sizes of the clusters 
did not allow us to trace reddening variations in any extinction map. Finally, we adopted the 
values obtained from the BH's maps, which are listed in Table 1. Their accuracy is 0.01 mag in 
$E(B-V)$ or 10$\%$ of the reddening, whichever is larger. Notice that a relative $C-T_1$ colour
shift smaller than $\sim$ 0.05 mag hardly turns into a meaningful difference when matching 
isochrones of the same metallicity to the cluster MSs. Likewise, the mean $C-T_1$ colour 
difference for isochrones of the same age and with the closest metallicity values that the 
Washington photometric system is able to distinguish ($\Delta$([Fe/H]) = 0.20 dex, \cite{getal97}) 
results larger than $\sim$ 0.20 mag. Consequently, the reddening uncertainties do not affect neither
the age nor the metallicity adopted for the clusters.

The LMC is located at a distance of 50 kpc \citep{ss10} and according to \cite{ss09} it has an 
average depth of (3.44 $\pm$ 1.16) kpc, which implies a difference in its distance modulus
as large as $\Delta$($(m-M)_o$) $\sim$ 0.30 mag. Given that the clusters could be placed in from of, or
behind the LMC, the simple assumption that all of them are at the same distance would lead to
age uncertainties of $\sigma$(log($t$)) $\le$ 0.06. However, the age difference of the isochrones
with the same metallicity bracketing the observed cluster MSs resulted to be in average 
$\Delta$(log($t$)) = 0.20. Therefore, we decided to adopt the value of the LMC distance modulus 
$(m - M)_o$ = 18.493 $\pm$ 0.008 reported by \citep{pietal13} for all the clusters. Indeed, we 
generally found an excellent match. 

As for the cluster metallicity we adopted the same value for all of them. \cite{pg13} showed that
during the last 3 Gyr (the age range of our cluster sample) the LMC cluster population has chemical 
evolved or spread within the range [Fe/H] = (-0.4 $\pm$ 0.2) dex. On the other hand, according to 
the Washington photometry metallicity sensitivity we should use isochrones for metallicity 
levels in steps of $\Delta$([Fe/H]) = 0.20 dex. Further higher metallicity resolution would lead to 
negligible changes in the isochrones overplotted on the cluster CMDs. However, we found that
only isochrones with the same age and metallicity differences larger than $\Delta$([Fe/H]) $\sim$
0.30-0.40 dex can be meaningfully differentiated due to the dispersion of the stars. For this
reasons, we adopted a mean value of [Fe/H] = -0.40 dex for cluster metallicities, and assumed
an error of $\sigma$([Fe/H]) = 0.20 dex.

In the matching procedure, which was performed looking at the cluster CMD, we used a subset of isochrones 
%ranging from $\Delta$(log($t$)) = -0.3 to +0.3 around the cluster age, 
and superimposed them on the cluster CMDs,
once they were properly shifted by the
corresponding $E(C-T_1)$ = 1.97$\times$$E(B-V)$ colour excesses and by the LMC apparent distance modulus
($M_{T_1}$ = $T_1$ + 0.58$E(B-V)$ - $(V-M_V)$) \citep{gs99}. Finally, we adopted as the cluster age 
the one corresponding to 
the isochrone which best reproduced the cluster main features in the CMD. The presence of RCs 
and/or Red Giant Branch stars in some cluster CMDs made the matching procedure easier. Table 1 
lists the resulting age estimates, while the bottom left panel in Fig. 2 (likewise the on-line
figures) show the corresponding isochrones superimposed. The observed dispersion 
seen in the cluster CMDs can be encompassed with a couple of isochrones bracketing the
derived mean age by $\Delta$(log($t$)) = $\pm$0.10. We also included these adjacent isochrones in 
Fig. 2 (bottom-left panel). 

Thirty-seven clusters in the sample have previous age estimates (see Table 1). \cite{petal13} made use 
of Washington $C,T_1$ photometry obtained at the CTIO Blanco telescope and the MOSAIC II 
camera and derived an age of log($t$) = 9.04$\pm$0.04 for HS\,156, in excellent agreement 
with our present value (9.00$\pm$0.10). We also found a tight agreement with \cite{pu00}, 
who derived from OGLE data an age of log($t$) = 8.2 for HS\,198 (8.15$\pm$0.10). 

A more careful analysis requires the comparison of our ages with those
of \citet[hereafter G10]{getal10} and \citet[hereafter P12]{poetal12} for 10 and 25 clusters 
in common, respectively. G10 used data from the Magellanic Clouds 
Photometric Survey \citep{zetal02} of clusters mostly distributed in the main body of 
the galaxy, which is highly crowded. Although they mention that field contamination is a 
severe effect in the extracted cluster CMDs and therefore influences the age estimates, 
no decontamination from field CMDs was carried out. It would not be unexpected
that some of the studied objects are not real star clusters, particularly
those with very uncertain age estimates [$\sigma$(log($t$)) $\ge$ 0.5]. Indeed, G10
estimated an age value for KMHK\,125 of  log($t$) = 7.2$\pm$0.3, for which our study 
suggests it could be a possible non-cluster (see. Fig. 4). This possibility alerts us 
to the fact that solely the circular extraction of the observed CMDs of clusters 
located in highly populated star fields is not enough neither for an accurate
isochrone fitting to the cluster MSs nor for confirming their physical nature.
\cite{zetal02} found little  visible evidence for incompleteness 
for $V$ $<$ 20 mag, corresponding to a MS turnoff of log($t$) $\approx$ 8.7. The present $CT_1$ data 
set nearly reaches the 100$\%$ completeness level at $T_1$ $\sim$ 22.0 mag, which corresponds to MS 
turnoffs of log($t$) $\sim$ 9.6.  Furthermore, the scale of our images is 0.274$\arcsec$ pix$^{-1}$, while 
that of the Magellanic Cloud Photometric Survey is 0.7$\arcsec$ pix$^{-1}$, so that crowding effects 
at the centre of the objects is more important in their images.

P12 derived ages from integrated photometry for the clusters of \cite{hetal03}. 
As is well known, the integrated light brings information about the composite stellar population
distributed along the line-of sight. For clusters of small angular size projected towards relatively 
crowded fields, the field contamination and stochastic effects, in addition to less deep photometric
data could mislead the data analysis. Indeed, \cite{betal13} compared the ages coming from 
\cite{pu00}, G10, and P12 and then, due to the large uncertainties in ages coming from
integrated magnitudes and colours, assigned ages to the clusters giving the
highest priority to HST ages from \cite{mg03}, followed by  G10, the OGLE data, and finally
P12. The best-fitting relation between G10 and P12 over 293 clusters in 
common resulted in an slope of 1.35$\pm$0.35 and in a Pearson coefficient of 0.77. 
From this result \cite{betal13} concluded that  there exists a good agreement between G10 and P12.

Fig. 5 shows the comparison between the ages derived by G10 (left panel) and P12 (right panel) and
our present values. The error bars correspond to the age uncertainties quoted by the authors,  while
the thick and thin lines represent the identity relationship and those shifted by 
$\pm$1$\sigma$(log($t$)$_{\rm our}$), respectively. Black filled squares represent clusters that
do not fullfill the requirement $\sigma$(log($t$)$_{\rm our}$) + $\sigma$(log($t$)$_{\rm pub}$)
$\ge$ $|$log($t$)$_{\rm our}$ - log($t$)$_{\rm pub}$ $|$. As can be seen, there is a reasonable
agreement, although several clusters significantly depart from the $\pm$1$\sigma$ strip. We have
carefully checked our multi-panel Fig. 2 (on-line material) in order to seek a possible explanation 
for such deviations and found out that stochastic effects caused by the presence of isolated bright 
stars, a significant field contamination, and a less faint magnitude limit  could be some
of the factors that affected the integrated light, and hence the P12's age estimates. Particularly,
the presence of a young field MS could cause the  P12's age for H88-270 and HS\,228 resulted younger
than our values. Conversely, an old field population could cause BSDL\,194, H88-240, and 
SL\,390 appear older at their integrated light. Examples of the presence of isolated bright
stars can be BSDL\,87, H88-238, 276, among others. We conclude that the present results point the
need of a better study of the LMC clusters with, e.g.,  8-m class telescopes.

\section{Discussion}

Our results suggest that  nearly 2/3 of the studied candidate star clusters would appear to 
be genuine physical systems. We also show that the ages derived by G10 and P12 can reflect those of the 
composite stellar populations of the LMC field. As far as we are aware, this is the first time that 
evidence is presented showing that some LMC candidate star clusters are not possible
genuine physical systems.

In order to examine whether there exists any dependence of the fraction of possible non-clusters with 
the position in the LMC, we have made use of the central deprojected galactocentric distances of the nine
Mosaic II fields computed by assuming that they are part of a disc having an inclination 
$i$ = 35.8$\degr$ and a position angle of the line of nodes of $\Theta$ = 145$\degr$ \citep{os02}. 
We refer the reader to Table 1 of \citet{ss10} which includes a summary of orientation measurements 
of the LMC disc plane, as well as their analysis of the orientation and other LMC disc quantities, 
supporting the present adopted values. Notice that most of the clusters catalogued by B08 located within
the Mosaic fields have already been studied previously -we have made use of them here-
\citep{getal10,pg13} and in the present work, so that there is no incompleteness effects in the estimated 
fraction of possible non-clusters in these areas.

Fig. 6 
illustrates the behaviour of the proportion of possible non-clusters as a function of the deprojected
distance. As can be seen, there exists
no visible trend but an averaged fraction of 0.13$\pm$0.06 within the LMC inner disc 
\citep[distance from the LMC centre $\le$ 4$\degr$,][]{betal98}.  This does not seem to be a 
significant percentage of the
catalogued clusters. If we assume a similar percentage for the whole catalogued LMC clusters 
distributed throughout the inner disc, we would expect that some $\sim$ (300$\pm$140) objects 
were not real stellar aggregates. In the case that the possible non-clusters had their own spatial 
distribution, it would be possible to recover the expected spatial distribution of genuine star 
clusters by using the former to correct that of the catalogued clusters.

We finally built the cluster frequencies (CFs) -the number of clusters per time unit as a function of age
\citep{betal13} - for the nine studied LMC regions with the aim of
investigating their variations in terms of the position in the galaxy recently reported by
\citet[hereafter P14]{p14}. Particularly he found that  30 Doradus turns out to be the region with the
highest relative frequency of the youngest clusters, while the log($t$) = 9-9.5 (1-3 Gyr) age
range is characterised by cluster formation at a higher rate in the inner regions 
 \cite[e.g. Bar, outer Bar; see][]{hz09} than in the
outer ones (e.g., Blue Arm, Constellation III, South-east arm). 

When building the CFs we took into account the unavoidable complication 
 that the cluster ages have associated uncertainties. Indeed, by considering such errors, 
the interpretation of the resulting CFs can differ appreciably from those obtained using only 
the measured ages without accounting for their errors. However, the treatment of
age errors in the CF is not a straightforward task. This happens
when an age value does not fall in the bin centre and, due to its errors,  has 
the chance to fall outside it. Note that, since we chose bin dimensions as large as the involved 
errors, such points should not fall on average far beyond the adjacent bins. However, this does not
 necessarily happen for all the age values, and we should consider at the same time any other possibility. 
We followed the procedure outlined by P14, which achieves a compromise between the age bin 
size and the age errors. Particularly, we considered the possibility that the extension 
covered by an age value (properly a segment of 2$\times$$\sigma$(age) long) may have a dimension smaller, 
similar or larger than the age bin wherein it is placed. The assigned weight  was computed as 
the fraction of its age segment that falls in the age bin. 

Fig. 7 shows the resultant CFs  built from the catalogue of cluster ages compiled by P14 
and from the ages estimated in this work, which encompass 90$\%$ of the whole sample of
catalogued clusters located in the studied LMC regions.  The CFs have been normalized 
to the total number of clusters used. We have used the deprojected distances 
of Fig. 6 as the 
independent spatial variable. At first glance, we confirm the existence of spatial 
variations of the LMC CF throughout the inner disc. However, we do not rule out that some percentage 
of such a variation comes from the fact that we are dealing with relatively small areas and consequently 
some age regimes result not well sampled. Nevertheless, 
Fig. 7 suggests that the innermost field contains a handful of clusters older than $\sim$ 2 Gyr;
the epoch of  an important burst of cluster formation that took place after a cluster age gap 
\citep{p11a,petal02}. It also shows that there has been a wider spread between different
CFs during the most recent 50 Myr of the galaxy lifetime.

\section{Conclusions}

The astrophysical properties of LMC star clusters have been the 
starting point to address galactic global issues such as the age-metallicity 
relationship and the cluster formation rate, among others. However, the number of clusters with
estimated properties is by far less than a half of the catalogued clusters.
Therefore, it results of great importance to enlarge the number of objects confirmed as 
genuine star clusters and to estimate their fundamental parameters.

We have used Washington photometry for 90 star cluster candidates of small angular size, 
typically $\sim$ 11$\arcsec$ in radius, distributed within nine fields (36$\times$36 arcmin$^2$) 
located in the LMC inner disc to
estimate their ages and to build their local cluster frequencies.  Some
clusters are projected towards relatively crowded fields. 

In order to assure a meaninful interpretation of the observed cluster CMDs, we applied a 
subtraction procedure to statistically clean them from the field star contamination. Thus, we 
could disentangle cluster features from those belonging to their surrounding fields.
The employed technique makes use of variable cells in order to reproduce the field CMD as 
closely as possible. 

Out of the 90 cluster candidates studied, 61 resulted to be genuine physical systems, whereas
the remaining ones were classified as possible non-clusters since either their CMDs and/or
the distribution of stars in the respective fields do not resemble those of stellar
aggregates. We statistically show that $\sim$ (13 $\pm$ 6)$\%$ of the catalogued clusters 
distributed within the inner disc (distance from the LMC $\le$ 4$\degr$) could be possible
non-clusters, independently of their deprojected distances. 

We derived the ages for the confirmed clusters from the fit of theoretical
isochrones computed for the Washingotn system to the cleaned cluster CMDs. When adjusting a 
subset of isochrones we took into account the LMC distance modulus and the individual star 
cluster colour excesses. The derived ages resulted to be in the age range 7.8 $\le$ log($t$)
$\le$ 9.2. Finally, we built CFs from ages available in the 
literature and from the ages estimated in this work, which encompass 90$\%$ of the whole 
sample of catalogued clusters located in the studied LMC regions. We found that
there exists some spatial variation of the LMC CF throughout the inner disc. 
Particularly, the innermost field (deprojected distance $\sim$ 0.56$\degr$) contains a handful of clusters 
older than $\sim$ 2 Gyr, while the wider spread between different
CFs has taken place during the most recent 50 Myr of the galaxy lifetime.

\section*{Acknowledgements}

We are grateful for the comments and suggestions raised by the anonymous
referee which helped us to improve the manuscript.
This research has made use of the SIMBAD database,
operated at CDS, Strasbourg, France, and draws upon data as distributed by the 
NOAO Science Archive. NOAO is operated by the Association of Universities for 
Research in Astronomy (AURA), Inc. under a cooperative agreement with the National 
Science Foundation.  
This work was partially supported by the Argentinian institutions
CONICET and Agencia Nacional de Promoci\'on Cient\'{\i}fica y Tecnol\'ogica (ANPCyT). 

%to be commented before sending to editor
%\bibliographystyle{mn2e_new} %style mn.bst
%\bibliography{paper} % your references file.bib

\begin{thebibliography}{43}
\expandafter\ifx\csname natexlab\endcsname\relax\def\natexlab#1{#1}\fi

\bibitem[{{Baumgardt} {et~al}\mbox{.}(2013){Baumgardt}, {Parmentier}, {Anders},
  \& {Grebel}}]{betal13}
{Baumgardt} H., {Parmentier} G., {Anders} P., {Grebel} E.~K., 2013, \mnras,
  430, 676

\bibitem[{{Bica} {et~al}\mbox{.}(2008){Bica}, {Bonatto}, {Dutra}, \&
  {Santos}}]{betal08}
{Bica} E., {Bonatto} C., {Dutra} C.~M., {Santos} J.~F.~C., 2008, \mnras, 389,
  678

\bibitem[{{Bica} {et~al}\mbox{.}(1998){Bica}, {Geisler}, {Dottori},
  {Clari{\'a}}, {Piatti}, \& {Santos}}]{betal98}
{Bica} E., {Geisler} D., {Dottori} H., {Clari{\'a}} J.~J., {Piatti} A.~E.,
  {Santos}, Jr. J.~F.~C., 1998, \aj, 116, 723

\bibitem[{{Brocato} {et~al}\mbox{.}(1996){Brocato}, {Castellani}, {Ferraro},
  {Piersimoni}, \& {Testa}}]{betal96}
{Brocato} E., {Castellani} V., {Ferraro} F.~R., {Piersimoni} A.~M., {Testa} V.,
  1996, \mnras, 282, 614

\bibitem[{{Burstein} \& {Heiles}(1982)}]{bh82}
{Burstein} D., {Heiles} C., 1982, \aj, 87, 1165

\bibitem[{{Canterna}(1976)}]{c76}
{Canterna} R., 1976, \aj, 81, 228

\bibitem[{{Choudhury}, {Subramaniam} \& {Piatti}(2014){Choudhury},
  {Subramaniam}, \& {Piatti}}]{chetal14}
{Choudhury} S., {Subramaniam} A., {Piatti} A., 2014, \mnras

\bibitem[{{de Grijs}, {Goodwin} \& {Anders}(2013){de Grijs}, {Goodwin}, \&
  {Anders}}]{dgetal13}
{de Grijs} R., {Goodwin} S.~P., {Anders} P., 2013, \mnras, 436, 136

\bibitem[{{Geisler} {et~al}\mbox{.}(1997){Geisler}, {Bica}, {Dottori},
  {Claria}, {Piatti}, \& {Santos}}]{getal97}
{Geisler} D., {Bica} E., {Dottori} H., {Claria} J.~J., {Piatti} A.~E.,
  {Santos}, Jr. J.~F.~C., 1997, \aj, 114, 1920

\bibitem[{{Geisler} {et~al}\mbox{.}(2003){Geisler}, {Piatti}, {Bica}, \&
  {Clari{\'a}}}]{getal03}
{Geisler} D., {Piatti} A.~E., {Bica} E., {Clari{\'a}} J.~J., 2003, \mnras, 341,
  771

\bibitem[{{Geisler} \& {Sarajedini}(1999)}]{gs99}
{Geisler} D., {Sarajedini} A., 1999, \aj, 117, 308

\bibitem[{{Glatt}, {Grebel} \& {Koch}(2010){Glatt}, {Grebel}, \&
  {Koch}}]{getal10}
{Glatt} K., {Grebel} E.~K., {Koch} A., 2010, {Ages and luminosities of young
  SMC/LMC star clusters and the recent star formation history of the Clouds}

\bibitem[{{Harris} \& {Zaritsky}(2009)}]{hz09}
{Harris} J., {Zaritsky} D., 2009, \aj, 138, 1243

\bibitem[{{Haschke}, {Grebel} \& {Duffau}(2011){Haschke}, {Grebel}, \&
  {Duffau}}]{hetal11}
{Haschke} R., {Grebel} E.~K., {Duffau} S., 2011, \aj, 141, 158

\bibitem[{{Hunter} {et~al}\mbox{.}(2003){Hunter}, {Elmegreen}, {Dupuy}, \&
  {Mortonson}}]{hetal03}
{Hunter} D.~A., {Elmegreen} B.~G., {Dupuy} T.~J., {Mortonson} M., 2003, \aj,
  126, 1836

\bibitem[{{Mackey} \& {Gilmore}(2003)}]{mg03}
{Mackey} A.~D., {Gilmore} G.~F., 2003, \mnras, 338, 85

\bibitem[{{Maia}, {Piatti} \& {Santos}(2014){Maia}, {Piatti}, \&
  {Santos}}]{metal14}
{Maia} F.~F.~S., {Piatti} A.~E., {Santos} J.~F.~C., 2014, \mnras, 437, 2005

\bibitem[{{Marigo} {et~al}\mbox{.}(2008){Marigo}, {Girardi}, {Bressan},
  {Groenewegen}, {Silva}, \& {Granato}}]{metal08}
{Marigo} P., {Girardi} L., {Bressan} A., {Groenewegen} M.~A.~T., {Silva} L.,
  {Granato} G.~L., 2008, \aap, 482, 883

\bibitem[{{Olsen} \& {Salyk}(2002)}]{os02}
{Olsen} K.~A.~G., {Salyk} C., 2002, \aj, 124, 2045

\bibitem[{{Palma} {et~al}\mbox{.}(2013){Palma}, {Clari{\'a}}, {Geisler},
  {Piatti}, \& {Ahumada}}]{petal13}
{Palma} T., {Clari{\'a}} J.~J., {Geisler} D., {Piatti} A.~E., {Ahumada} A.~V.,
  2013, \aap, 555, A131

\bibitem[{{Piatti}(2011{\natexlab{a}})}]{p11c}
{Piatti} A.~E., 2011{\natexlab{a}}, \mnras, 416, L89

\bibitem[{{Piatti}(2011{\natexlab{b}})}]{p11a}
---, 2011{\natexlab{b}}, \mnras, 418, L40

\bibitem[{{Piatti}(2011{\natexlab{c}})}]{p11b}
---, 2011{\natexlab{c}}, \mnras, 418, L69

\bibitem[{{Piatti}(2012{\natexlab{a}})}]{p12b}
---, 2012{\natexlab{a}}, \mnras, 422, 1109

\bibitem[{{Piatti}(2012{\natexlab{b}})}]{p12a}
---, 2012{\natexlab{b}}, \aap, 540, A58

\bibitem[{{Piatti}(2014)}]{p14}
---, 2014, \mnras, 437, 1646

\bibitem[{{Piatti} \& {Bica}(2012)}]{pb12}
{Piatti} A.~E., {Bica} E., 2012, \mnras, 425, 3085

\bibitem[{{Piatti} {et~al}\mbox{.}(2003{\natexlab{a}}){Piatti}, {Bica},
  {Geisler}, \& {Clari{\'a}}}]{petal03b}
{Piatti} A.~E., {Bica} E., {Geisler} D., {Clari{\'a}} J.~J.,
  2003{\natexlab{a}}, \mnras, 344, 965

\bibitem[{{Piatti} {et~al}\mbox{.}(2011){Piatti}, {Clari{\'a}}, {Parisi}, \&
  {Ahumada}}]{petal11}
{Piatti} A.~E., {Clari{\'a}} J.~J., {Parisi} M.~C., {Ahumada} A.~V., 2011,
  \pasp, 123, 519

\bibitem[{{Piatti} \& {Geisler}(2013)}]{pg13}
{Piatti} A.~E., {Geisler} D., 2013, \aj, 145, 17

\bibitem[{{Piatti} {et~al}\mbox{.}(2003{\natexlab{b}}){Piatti}, {Geisler},
  {Bica}, \& {Clari{\'a}}}]{petal03a}
{Piatti} A.~E., {Geisler} D., {Bica} E., {Clari{\'a}} J.~J.,
  2003{\natexlab{b}}, \mnras, 343, 851

\bibitem[{{Piatti} {et~al}\mbox{.}(1999){Piatti}, {Geisler}, {Bica},
  {Clari{\'a}}, {Santos}, {Sarajedini}, \& {Dottori}}]{petal99}
{Piatti} A.~E., {Geisler} D., {Bica} E., {Clari{\'a}} J.~J., {Santos}, Jr.
  J.~F.~C., {Sarajedini} A., {Dottori} H., 1999, \aj, 118, 2865

\bibitem[{{Piatti}, {Geisler} \& {Mateluna}(2012){Piatti}, {Geisler}, \&
  {Mateluna}}]{petal12}
{Piatti} A.~E., {Geisler} D., {Mateluna} R., 2012, \aj, 144, 100

\bibitem[{{Piatti} {et~al}\mbox{.}(2009){Piatti}, {Geisler}, {Sarajedini}, \&
  {Gallart}}]{petal09}
{Piatti} A.~E., {Geisler} D., {Sarajedini} A., {Gallart} C., 2009, \aap, 501,
  585

\bibitem[{{Piatti} {et~al}\mbox{.}(2002){Piatti}, {Sarajedini}, {Geisler},
  {Bica}, \& {Clari{\'a}}}]{petal02}
{Piatti} A.~E., {Sarajedini} A., {Geisler} D., {Bica} E., {Clari{\'a}} J.~J.,
  2002, \mnras, 329, 556

\bibitem[{{Pietrzy{\'n}ski} {et~al}\mbox{.}(2013){Pietrzy{\'n}ski}, {Graczyk},
  {Gieren}, {Thompson}, {Pilecki}, {Udalski}, {Soszy{\'n}ski}, {Koz{\l}owski},
  {Konorski}, {Suchomska}, {Bono}, {Moroni}, {Villanova}, {Nardetto},
  {Bresolin}, {Kudritzki}, {Storm}, {Gallenne}, {Smolec}, {Minniti}, {Kubiak},
  {Szyma{\'n}ski}, {Poleski}, {Wyrzykowski}, {Ulaczyk}, {Pietrukowicz},
  {G{\'o}rski}, \& {Karczmarek}}]{pietal13}
{Pietrzy{\'n}ski} G. {et~al.}, 2013, \nat, 495, 76

\bibitem[{{Pietrzy{\'n}ski} \& {Udalski}(2000)}]{pu00}
{Pietrzy{\'n}ski} G., {Udalski} A., 2000, \actaa, 50, 337

\bibitem[{{Popescu}, {Hanson} \& {Elmegreen}(2012){Popescu}, {Hanson}, \&
  {Elmegreen}}]{poetal12}
{Popescu} B., {Hanson} M.~M., {Elmegreen} B.~G., 2012, \apj, 751, 122

\bibitem[{{Schlegel}, {Finkbeiner} \& {Davis}(1998){Schlegel}, {Finkbeiner}, \&
  {Davis}}]{setal98}
{Schlegel} D.~J., {Finkbeiner} D.~P., {Davis} M., 1998, \apj, 500, 525

\bibitem[{{Subramanian} \& {Subramaniam}(2009)}]{ss09}
{Subramanian} S., {Subramaniam} A., 2009, \aap, 496, 399

\bibitem[{{Subramanian} \& {Subramaniam}(2010)}]{ss10}
---, 2010, \aap, 520, A24

\bibitem[{{Udalski}(2003)}]{u03}
{Udalski} A., 2003, \actaa, 53, 291

\bibitem[{{Zaritsky} {et~al}\mbox{.}(2002){Zaritsky}, {Harris}, {Thompson},
  {Grebel}, \& {Massey}}]{zetal02}
{Zaritsky} D., {Harris} J., {Thompson} I.~B., {Grebel} E.~K., {Massey} P.,
  2002, \aj, 123, 855

\end{thebibliography}
% 
%to be uncommented before sending to editor

%

\clearpage

\begin{table}
\tiny
\caption{Published and present LMC cluster fundamental parameters.}
\begin{tabular}{@{}lcccccccc}\hline
ID         &   R.A.    & Dec. & {\it l}       & b       &    r            & E(B-V) &  log($t$/yr)  & Note \\ 
           & (h m s)   & ($\degr$ $\arcmin$ $\arcsec$) & ($\degr$)  & ($\degr$)  & ($\arcmin$)  & (mag)  &  & \\\hline

GKK-O222   &  05 09 00  & -67 59 00 &  278.704 & -34.537 &  0.14 &  0.055 &     9.20   &  \\
HS\,228      &  05 19 24  & -68 52 52 &  279.522 & -33.415 &  0.27 &  0.060 &     9.10   &  8.610$\pm$0.055 (4) \\ 
H88-270    &  05 18 47  & -69 16 37 &  279.999 & -33.392 &  0.18 &  0.090 &     9.10   &  7.950$\pm$0.150 (4)\\
HS\,156      &  05 11 11  & -67 37 37 &  278.227 & -34.414 &  0.18 &  0.061 &     9.00   &  9.04$\pm$0.04 (1) \\
SL\,390      &  05 19 54  & -68 57 53 &  279.609 & -33.354 &  0.32 &  0.081 &     9.00      & 9.200$\pm$0.030 (4) \\
BSDL\,1334   &  05 21 14  & -68 47 00 &  279.369 & -33.271 &  0.18 &  0.060 &     9.00      &  \\    
H88-259    &  05 17 20  & -69 09 25 &  279.890 & -33.542 &  0.18 &  0.081 &     8.90       &  \\
H88-276    &  05 19 55  & -68 48 07 &  279.418 & -33.384 &  0.23 &  0.060 &     8.90      &  8.180$\pm$0.215 (4)\\
GKK-O220   &  05 10 18  & -67 51 00 &  278.512 & -34.447 &  0.23 &  0.060 &     8.90       &  \\
OGLE\,264    &  05 15 21  & -69 06 27 &  279.875 & -33.725 &  0.18 &  0.078 &     8.80      &  \\
SL\,124e     &  04 56 31  & -69 58 54 &  281.410 & -35.091 &  0.14 &  0.115 &     8.70    & 8.730$\pm$0.170 (4)\\
SL\,124w     &  04 56 29  & -69 59 00 &  281.413 & -35.094 &  0.14 &  0.115 &     8.70      &  \\
H88-261    &  05 17 26  & -69 06 55 &  279.838 & -33.542 &  0.32 &  0.081 &     8.70      &  \\ 
HS\,200      &  05 15 36  & -69 08 21 &  279.907 & -33.697 &  0.18 &  0.086 &     8.70   & 8.250$\pm$0.800 (4) \\
OGLE\,122    &  05 07 19  & -68 20 55 &  279.179 & -34.605 &  0.18 &  0.059 &     8.60      &  \\ 
H88-260    &  05 17 20  & -69 12 49 &  279.956 & -33.530 &  0.18 &  0.090 &     8.60        &  \\
OGLE\,271    &  05 15 39  & -68 54 31 &  279.635 & -33.741 &  0.27 &  0.078 &     8.60        &  \\
H88-285    &  05 21 03  & -69 05 51 &  279.741 & -33.229 &  0.18 &  0.081 &     8.50        & 8.070$\pm$0.025 (4)\\
H88-326    &  05 43 29  & -69 09 44 &  279.484 & -31.242 &  0.18 &  0.072 &     8.50   & 8.45$\pm$0.3 (2) \\
H88-249    &  05 16 31  & -69 10 58 &  279.939 & -33.608 &  0.14 &  0.090 &     8.50       &  \\
H88-286    &  05 21 07  & -69 08 09 &  279.787 & -33.216 &  0.14 &  0.081 &     8.50  & 8.45$\pm$0.5 (2) \\
H88-272    &  05 19 05  & -68 52 14 &  279.515 & -33.445 &  0.23 &  0.060 &     8.50           &  \\
H88-252    &  05 16 43  & -69 12 13 &  279.958 & -33.586 &  0.23 &  0.090 &     8.40           & 8.470$\pm$0.255 (4) \\
H88-34     &  04 55 39  & -67 49 19 &  278.900 & -35.793 &  0.14 &  0.058 &     8.40            & 8.250$\pm$0.110 (4) \\    
H88-32     &  04 55 39  & -67 43 34 &  278.788 & -35.819 &  0.14 &  0.058 &     8.40           &  8.150$\pm$0.155 (4)\\
BSDL\,194    &  04 54 05  & -69 45 30 &  281.227 & -35.359 &  0.14 &  0.102 &     8.40                 & 8.640$\pm$0.135 (4) \\
H88-325    &  05 43 15  & -69 02 03 &  279.337 & -31.275 &  0.18 &  0.072 &     8.35          &  \\
HS\,113      &  05 05 54  & -68 14 15 &  279.085 & -34.759 &  0.23 &  0.059 &     8.30           &  \\
H88-253    &  05 16 50  & -69 03 35 &  279.786 & -33.605 &  0.18 &  0.081 &     8.30           &  \\
H88-240    &  05 16 04  & -69 06 09 &  279.853 & -33.664 &  0.27 &  0.081 &     8.30          & 8.650$\pm$0.050 (4)\\ 
OGLE\,346    &  05 19 09  & -69 15 36 &  279.972 & -33.363 &  0.25 &  0.090 &     8.20          &  \\ 
H88-281    &  05 20 21  & -69 14 48 &  279.932 & -33.262 &  0.23 &  0.090 &     8.20          & 8.270$\pm$0.215 (4)\\ 
BRHT\,62a    &  05 00 05  & -67 48 03 &  278.736 & -35.397 &  0.32 &  0.062 &     8.20   & 8.9$\pm$0.5 (2) \\
BSDL\,1364   &  05 21 46  & -68 43 53 &  279.298 & -33.233 &  0.09 &  0.060 &     8.20             &  \\ 
H88-287    &  05 21 09  & -69 07 02 &  279.763 & -33.216 &  0.23 &  0.081 &     8.20              &  8.310$\pm$0.085 (4) \\ 
SL\,408A     &  05 21 05  & -69 04 16 &  278.896 & -34.889 &  0.23 &  0.055 &     8.15   & 7.7$\pm$0.3 (2) \\
HS\,198      &  05 15 26  & -69 03 02 &  279.807 & -33.730 &  0.14 &  0.078 &     8.15   & 8.2 (3)  \\
H88-313    &  05 41 21  & -69 03 46 &  279.391 & -31.441 &  0.18 &  0.064 &     8.10           &  7.930$\pm$0.085 (4)\\
H88-232    &  05 15 22  & -69 02 32 &  279.798 & -33.737 &  0.18 &  0.078 &     8.10             & 8.140$\pm$0.095 (4)\\ 
H88-306    &  05 40 24  & -69 15 10 &  279.623 & -31.505 &  0.14 &  0.063 &     8.10              &  \\ 
H88-238    &  05 15 47  & -69 14 39 &  280.027 & -33.659 &  0.23 &  0.090 &     8.10                & 7.950$\pm$0.050 (4) \\ 
KMHK\,1029   &  05 31 57  & -67 52 43 &  278.122 & -32.435 &  0.27 &  0.058 &     8.00                & 7.920$\pm$0.055 (4)\\
BSDL\,192    &  04 54 05  & -69 40.90 &  281.138 & -35.382 &  0.18 &  0.102 &     8.00  & 8.3$\pm$0.3 (2) \\
BSDL\,320    &  04 57 08  & -70 06 42 &  281.542 & -35.002 &  0.14 &  0.115 &     8.00              &  8.300$\pm$0.200 (4) \\
KMHK\,335    &  04 56 51  & -70 06 03 &  281.537 & -35.029 &  0.09 &  0.115 &     8.00                  & 8.390$\pm$0.060 (4)\\
HS\,205      &  05 16 32  & -68 55 07 &  279.627 & -33.660 &  0.27 &  0.081 &     8.00                 &  \\
OGLE\,363    &  05 20 04  & -69 15 55 &  279.959 & -33.282 &  0.18 &  0.090 &     8.00                  &  \\
OGLE\,127    &  05 07 32  & -67 34 13 &  278.252 & -34.766 &  0.23 &  0.061 &     8.00                 &  \\
KMHK\,993    &  05 30 34  & -68 09 27 &  278.469 & -32.526 &  0.18 &  0.062 &     8.00                 &  \\
H88-321    &  05 42 08  & -69 22 00 &  279.737 & -31.341 &  0.18 &  0.063 &     7.90                & 7.570$\pm$0.090 (4)\\ 
BRHT\,60b    &  04 56 25  & -67 56 21 &  279.014 & -35.691 &  0.14 &  0.060 &     7.90 &  7.6$\pm$0.5 (2) \\
OGLE\,297    &  05 16 52  & -69 04 13 &  279.798 & -33.600 &  0.27 &  0.081 &     7.90           & 7.610$\pm$0.235 (4)\\ 
BSDL\,87     &  04 50 58  & -67 36 35 &  278.808 & -36.279 &  0.18 &  0.050 &     7.90     & 7.500$\pm$0.075 (4) \\
BRHT\,45b    &  04 56 52  & -68 00 20 &  279.078 & -35.632 &  0.14 &  0.060 &     7.90  & 7.6$\pm$0.5 (2) \\
NGC\,1764    &  04 56 28  & -67 41 41 &  278.724 & -35.753 &  0.23 &  0.058 &     7.90  & 7.8$\pm$0.3 (2) \\
H88-315      &  05 41 38  & -69 18 48 &  279.680 & -31.391 &  0.18 &  0.063 &     7.90  & 7.4$\pm$0.5 (2) \\
HS\,41       &  04 51 30  & -67 27 15 &  278.605 & -36.276 &  0.18 &  0.048 &     7.80  & 7.9$\pm$0.3 (2) \\
H88-329      &  05 43 43  & -69 13 23 &  279.554 & -31.216 &  0.18 &  0.072 &     7.80          &  7.210$\pm$0.110 (4)\\
NGC\,1885    &  05 15 07  & -68 58 43 &  279.729 & -33.772 &  0.32 &  0.081 &     7.80                 & 8.300$\pm$0.030 (4)\\ 
OGLE\,340    &  05 18 47  & -69 13 32 &  279.939 & -33.402 &  0.36 &  0.090 &     7.60                 &  \\
H88-327    &  05 43 38  & -69 15 51 &  279.603 & -31.219 &  0.27 &  0.074 &     7.50                  &  \\
\hline
\end{tabular}

\medskip
\noindent Note: (1) Palma et al. (2013); (2) Glatt et al. (2010); (3) Pietrzy{\'n}ski \& Udalski (2000);
(4) Popescu et al. (2014).
\end{table}

\clearpage

\begin{figure}
\centerline{\psfig{figure=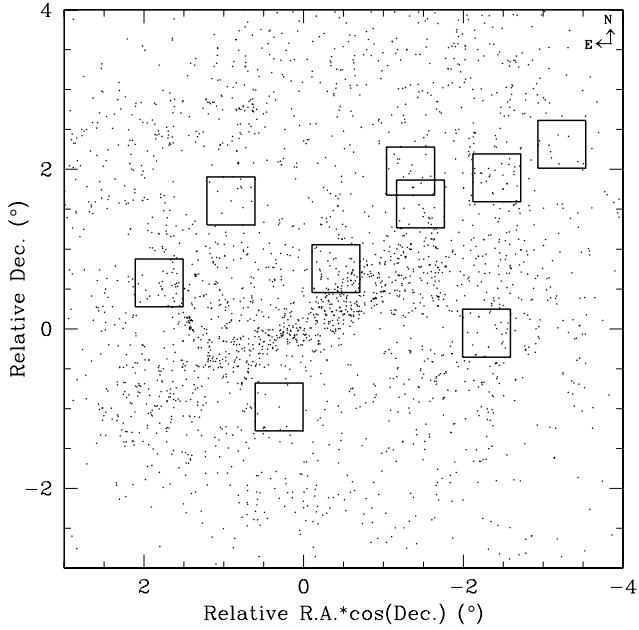,width=84mm}}
\caption{Distribution of the LMC star cluster candidates catalogued by Bica et al. (2008, dots)
and the Mosaic II fields (boxes) wherein the studied objects are located.}
\label{fig1}
\end{figure}

\begin{figure}
\centerline{\psfig{figure=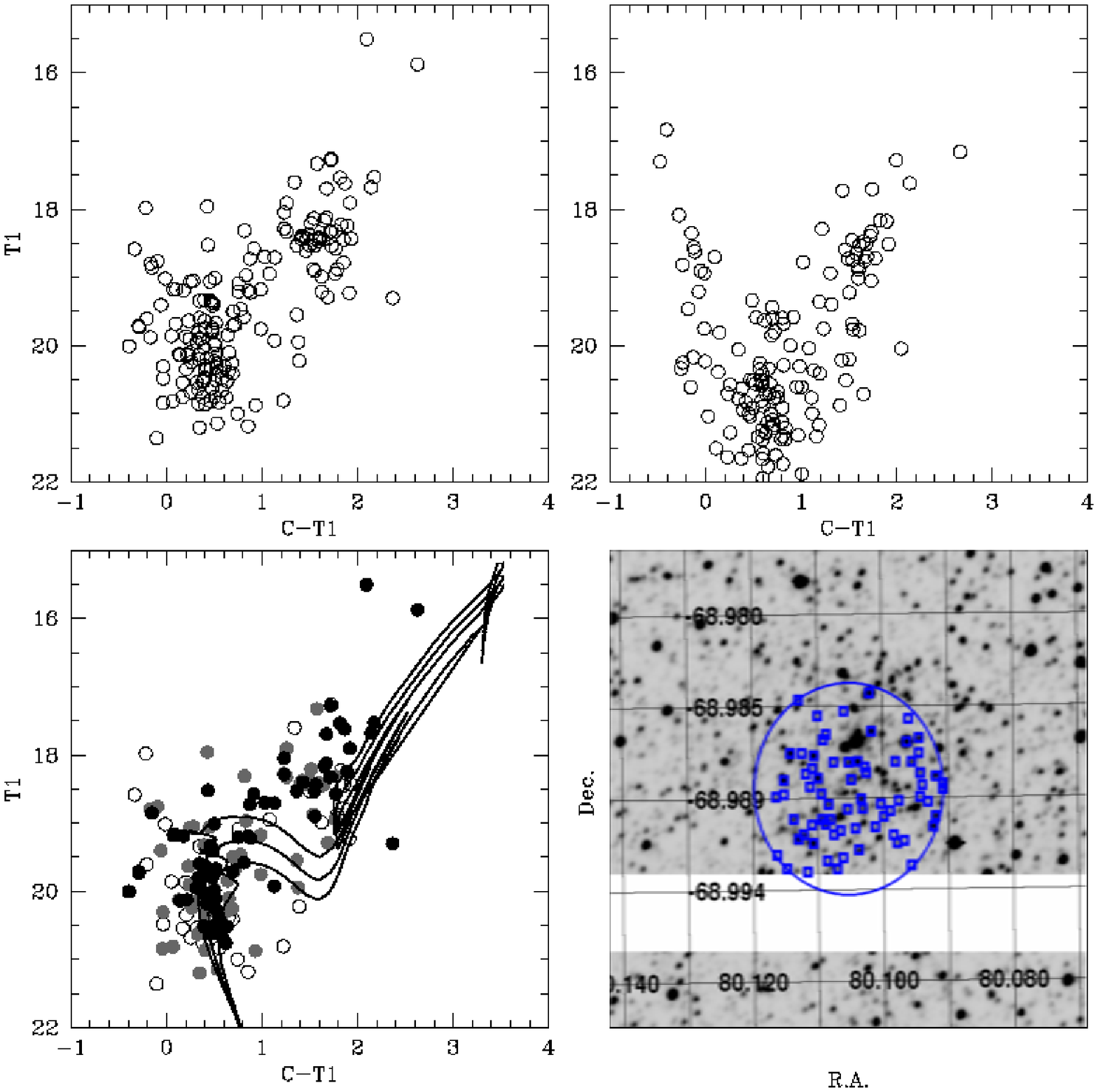,width=84mm}}
\caption{CMDs for stars in the field of SL\,390: the observed CMD for the stars distributed 
within the cluster radius (upper-left panel); a field CMD for an annulus centred on the 
cluster, with an internal radius 3 times the cluster radius and an area equal cluster area
(upper-right panel); the cleaned cluster CMD (bottom-left). Colour-scaled symbols represent 
stars that statistically belong to the field (P $\le$ 25\%,  white), stars that might 
belong either to the field or to the cluster (P  $=$ 50\%, gray), and stars that 
predominantly populate the cluster region (P $\ge$ 75\%,  black). Three isochrones from Marigo et 
al. (2008) for log($t$) (see Table 1) and  log($t$) $\pm$ 0.10 are also superimposed. An 
enlargement of the $T_1$ image centred on the cluster with a circle representing the adopted cluster radius
and stars with P $\ge$ 75$\%$ marked is shown in the 
bottom-right panel. North is upwards, and East is to the left.}
\label{fig2}
\end{figure}

\begin{figure}
\centerline{\psfig{figure=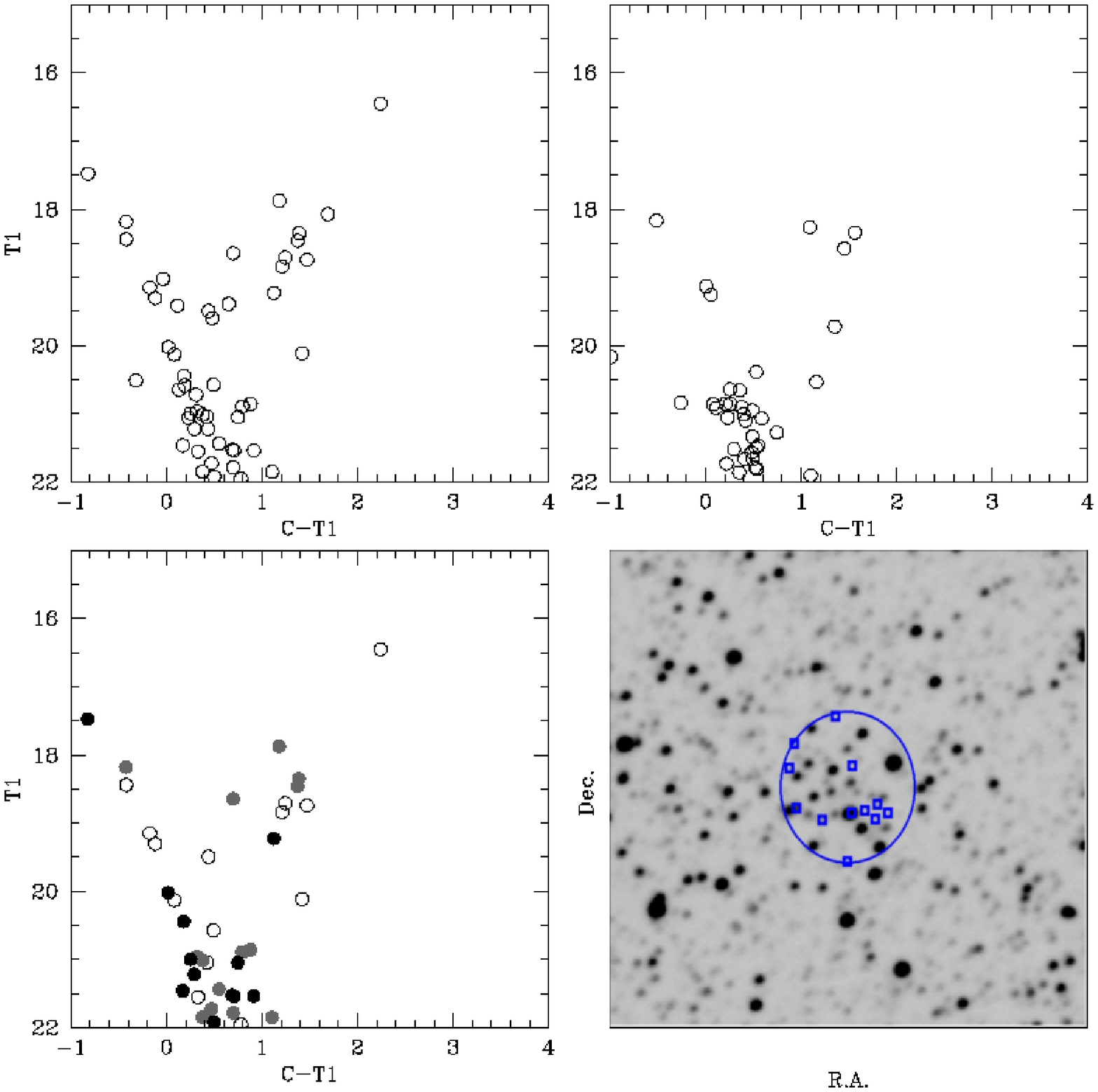,width=84mm}}
\caption{CMDs for stars in the field of BSDL\,616: the observed CMD for the stars distributed 
within the cluster radius (upper-left panel); a field CMD for an annulus centred on the 
cluster, with an internal radius 3 times the cluster radius and an area equal cluster area
(upper-right panel); the cleaned cluster CMD (bottom-left). Colour-scaled symbols represent 
stars that statistically belong to the field (P $\le$ 25\%, white), stars that might 
belong either to the field or to the cluster (P  $=$ 50\%, gray), and stars that 
predominantly populate the cluster region (P $\ge$ 75\%, black). An enlargement of the $T_1$ 
image centred on the cluster with a circle representing the adopted cluster radius
and stars with P $\ge$ 75$\%$ marked is shown in the 
bottom-right panel. North is upwards, and East is to the left.}
\label{fig3}
\end{figure}

\begin{figure}
\centerline{\psfig{figure=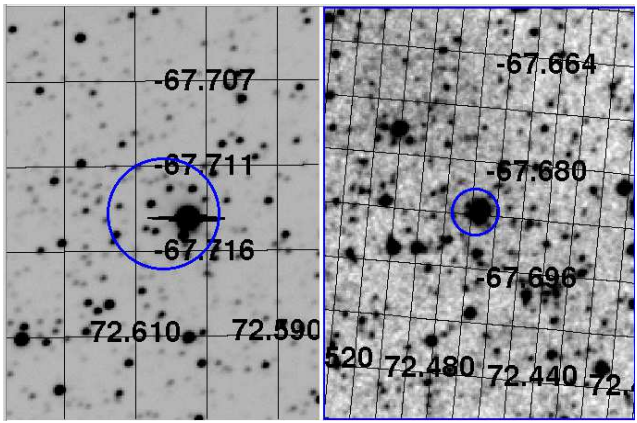,width=84mm}}
\caption{An enlargement of the $T_1$ image centred on KMHK\,125 (left panel) and that from
the DSS image (right panel). North is up and east is to the left. Notice that the
$T_1$ image is set on the telescope coordinates.}
\label{fig4}
\end{figure}

\begin{figure}
\centerline{\psfig{figure=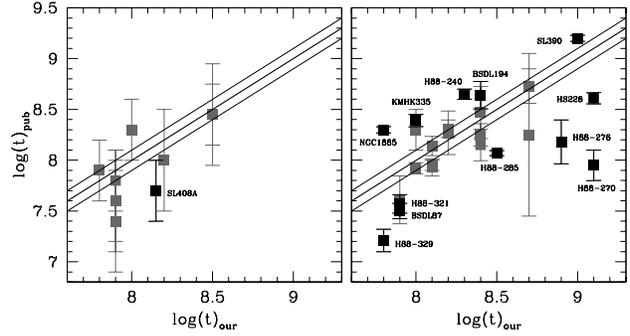,width=84mm}}
\caption{Difference between G10 (left panel) and P12 (right panel) and present age values. 
The error bars refer to age uncertainties quoted by the authors, and the thick and thin lines 
represent the identity relationship and those shifted by $\pm$1$\sigma$(log($t$)$_{\rm our}$), 
respectively. Black filled squares represent clusters that do not fullfill the requirement 
$\sigma$(log($t$)$_{\rm our}$) + $\sigma$(log($t$)$_{\rm pub}$)
$\ge$ $|$log($t$)$_{\rm our}$ - log($t$)$_{\rm pub}$ $|$.}
\label{fig5}
\end{figure}

\begin{figure}
\centerline{\psfig{figure=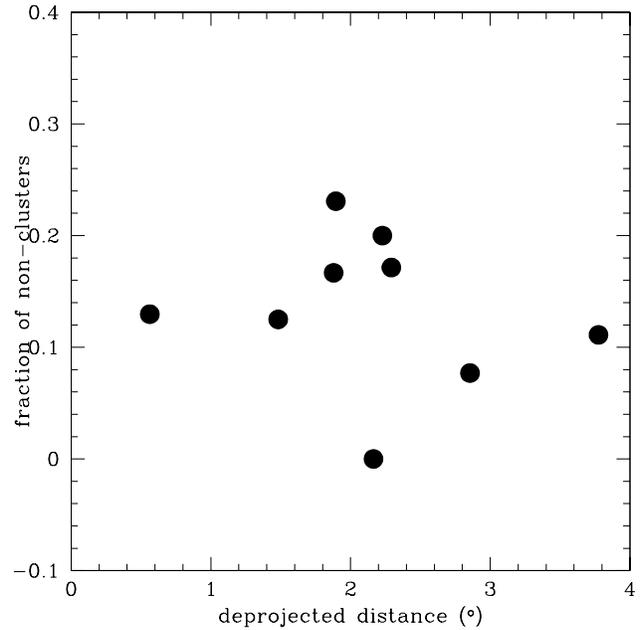,width=84mm}}
\caption{Fraction of possible non-clusters as a function of the deprojected distance (see text
for details).}
\label{fig6}
\end{figure}

\begin{figure}
\centerline{\psfig{figure=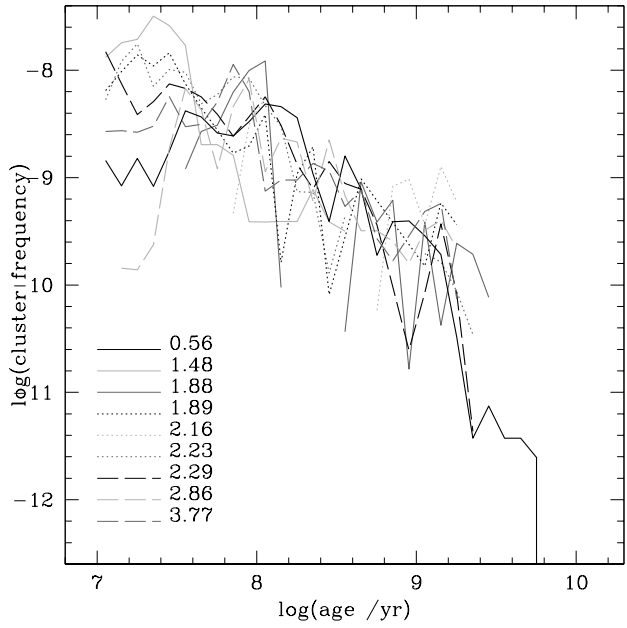,width=84mm}}
\caption{Cluster frequencies for different LMC fields. The deprojected distances 
($\degr$) of each field are indicated in the bottom-left corner.}
\label{fig6}
\end{figure}

\label{lastpage}

\end{document}